\documentclass{cpcauth} 
\usepackage{epsfig}

\usepackage{amssymb}
\usepackage{graphicx}
\journal{Computer Physics Communications}
\begin{document}
\begin{frontmatter}

\title{Crossover effects in the bond-diluted Ising model in three dimensions}

\author[Rouen]{P.E. Berche},
\author[Leipzig]{C. Chatelain},
\author[Nancy]{B. Berche} and 
\author[Leipzig]{W. Janke}

\address[Rouen]{Groupe de Physique des Mat\'eriaux, 
	Universit\'e de Rouen,
	F-76821 Mont Saint-Aignan Cedex, France}
\address[Leipzig]{Institut f\"ur Theoretische Physik, Universit\"at Leipzig, 
	Augustusplatz 10/11, D-04109 Leipzig, Germany}
\address[Nancy]{Laboratoire de Physique des Mat\'eriaux,
	Universit\'e Henri Poincar\'e, 
    BP 239,
	F-54506 Vandoeuvre les Nancy Cedex, France}

\begin{abstract}
We investigate by Monte Carlo simulations the critical properties of the 
three-dimensional bond-diluted Ising model. The phase diagram is determined 
by locating the maxima of the magnetic susceptibility and is compared to 
mean-field and effective-medium approximations. The calculation of the 
size-dependent effective critical exponents shows the competition between 
the different fixed points of the model as a function of the bond dilution.
\end{abstract} 

\end{frontmatter}

\section{Introduction}

The qualitative influence of quenched disorder at second-order phase transitions is well
understood since Harris proposed a relevance criterion~\cite{Harris74} based on the knowledge
of the specific heat critical exponent $\alpha_{\rm{pure}}$ of the pure model: when $\alpha_{\rm{pure}}$ is positive,
the disordered system will reach a new fixed point with new critical exponents whereas if
$\alpha_{\rm{pure}}$ is negative, the same universality class will persist.

As a paradigmatic model, the three-dimensional (3D) disordered Ising model 
characterized by $\alpha_{\rm{pure}}=0.109(4)$ 
has been extensively studied by:

$\bullet$ Renormalization group methods in the weak quenched dilution regime. 
The best estimates for
the critical exponents obtained by this method are~\cite{Pelissetto2000}: 

$\nu=0.678\ \pm\ 0.010,$

$\eta=0.030\ \pm\ 0.003,$

$\gamma=1.330\ \pm\  0.017$.

$\bullet$ Monte Carlo simulations of the site-diluted case for which the
following exponents have been found~\cite{Ballesteros1998}: 

$\nu=0.6837\ \pm\ 0.0053,$

$\beta=0.3546\ \pm\ 0.0028,$

$\gamma=1.342\ \pm\  0.010$.

$\bullet$ Experimental investigations.

\noindent For a review of these different results, see Ref.~\cite{Folk2001}.

The general picture which is now widely accepted is that, starting from the pure Ising model
(with exponents $\nu=0.6304\ \pm\ 0.0013$, $\gamma=1.2396\ \pm \ 0.0013$,
$\ \beta=0.3258\ \pm \ 0.0014$, see Ref.~\cite{Guida1998}),
the critical temperature decreases and eventually vanishes at the percolation threshold below
which there is no longer any long-range order in the system, due to the absence of percolating
clusters. In the dilution-temperature plane, the critical line $T_c (p)$ which delimits
ferromagnetic and paramagnetic phases has two end-points (namely the pure system and the 
percolation point) described by unstable fixed points and it is commonly believed that the
quenched disordered system has universal properties described by a unique stable fixed point.
According to this picture, the competition between fixed points possibly leads to an effective
variation of universal critical quantities.

In this paper, we first determine the phase diagram of the bond-diluted problem (up to now,
only the site-diluted case has been studied) and compare it to a prediction in the single-bond
effective-medium approximation and then we try to illustrate the competition between fixed
points leading to crossover regimes in some physical quantities.

\section{Phase diagram}

The bond-diluted Ising model is defined by the following Hamiltonian with independent quenched
random interactions:

\begin{equation}
-\beta H=\sum_{(i,j)} K_{ij}\delta_{\sigma_i ,\sigma_j} \qquad (\sigma_i=\pm 1).
\label{eq.1}
\end{equation}

The coupling strengths are allowed to take two different values $K_{ij}=K \equiv J/k_BT$ and
$0$ with probabilities $p$ and $1-p$, respectively,

\begin{equation}
P(K_{ij})=p\delta(K_{ij}-K)+(1-p)\delta(K_{ij}),
\label{eq.2}
\end{equation}
$c=1-p$ being the concentration of missing bonds, which play the role of the non-magnetic
impurities.
The simulation technique is based on the Swendsen-Wang cluster algorithm with periodic boundary
conditions in the three space directions.

\begin{figure}
\epsfysize=5.5cm
\begin{center}
\mbox{\epsfbox{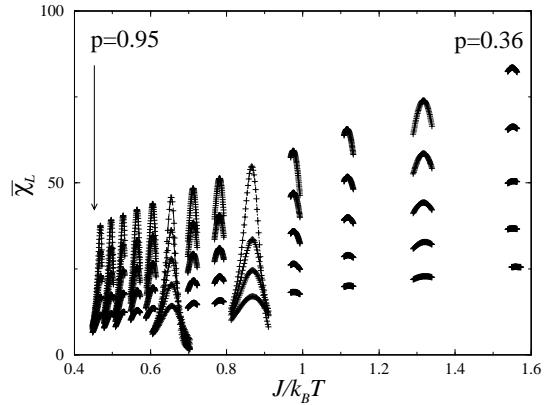}}
\end{center}\vskip 0cm
\caption{Variation of the average magnetic susceptibility $\bar\chi_L$ versus the coupling
strength $K=J/k_B T$ for several concentrations $p$ and $L=8,10,12,14,16,18,20$. For each value of $p$
and each size, only one value of $K$ has been simulated: the
extension of the $K$ values has been obtained by the standard histogram reweighting technique.}
\vspace*{-0.3cm}
\label{Fig1}
\end{figure}

The phase diagram is obtained numerically from the maxima of a diverging quantity (Fig.~\ref{Fig1}).
Here we
choose the susceptibility, since the stability of the disordered fixed point implies that the
specific-heat exponent is negative in the random system. Thus, the error in this quantity
is larger than for the susceptibility. The percolation threshold is located at
$p_c \approx 0.2488$.

\begin{figure}
\epsfysize=6.3cm
\begin{center}
\mbox{\epsfbox{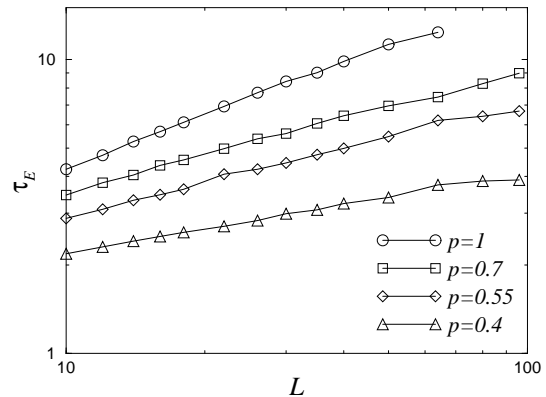}}
\end{center}\vskip 0cm
\caption{Variation on a log-log scale of the energy autocorrelation time 
$\tau_E$ versus the size $L$ of the system. The energy autocorrelation time
increases with the concentration of magnetic bonds $p$ and its greatest 
value obtained for $L=96,\ p=0.7$ is around $9$.}
\label{Fig2}
\end{figure}

To get an accurate determination of the maxima of the susceptibility, we used
the histogram
reweighting technique with $2\,500$ Monte Carlo sweeps (MCS) and between
$2\,500$ and $5\,000$ samples of disorder. The number of Monte Carlo sweeps
is justified by the increasing behaviour of the energy autocorrelation time
and we chose for each size at least $250$ independent measurements of the
physical quantities $(N_{\rm MCS} > 250\ \tau_E)$.

For a second-order phase transition, the autocorrelation time is expected to behave as
$L^z$ at the critical point where $z$ is the dynamical critical exponent
(Fig.~\ref{Fig2}). For the disordered Ising model, we get the values of $z$
shown in Table \ref{Tab1}.
We see that the critical slowing down weakens for the disordered model and
becomes smaller when the concentration of magnetic bonds $p$ decreases, but it
is necessary to increase the number of disorder realisations
when $p$ decreases because of the vicinity of the percolation threshold.

      \begin{table}
       \caption{The dynamical critical exponent $z$ as obtained from linear
fits of $\log \tau_E$ vs $\log L$.   \label{Tab1}}
       \vspace*{0.1cm}
        \begin{tabular}{|l|l|l|l|l|} \hline
        $\quad p\quad$  & $\quad 1\quad$  & $\quad 0.7\quad$ & $\quad 0.55\quad$ & 
$\quad 0.4\quad$ \\ \hline
        $\quad z\quad$  & $\quad 0.59\quad$ & $\quad 0.41\quad$  & $\quad 0.38\quad$  
& $\quad 0.27\quad$ \\ \hline
        \end{tabular}
        \end{table}

\begin{figure}
\epsfysize=5.5cm
\begin{center}
\mbox{\epsfbox{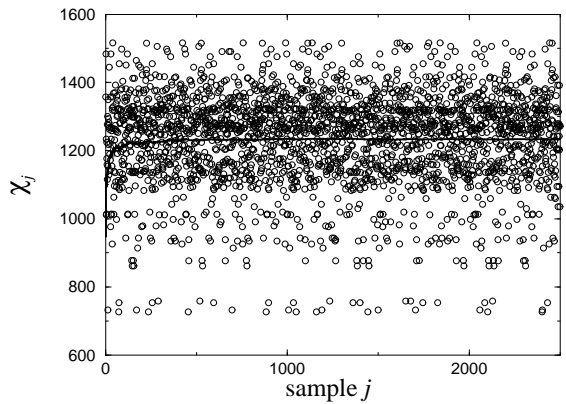}}
\end{center}\vskip 0cm
\caption{Distribution of the susceptibility for the different disorder 
realisations of the Ising model with $L=96$ and a concentration of magnetic 
bonds $p=0.7$. The average value over the samples $\bar\chi$ is shown by 
the black line.}
\vspace*{-0.5cm}
\label{Fig3}
\end{figure}

In order to check the quality of the averaging techniques, we can study the stability of
the susceptibility for the largest size considered versus the number of Monte Carlo sweeps
involved in the thermal average. The results are given in Table \ref{Tab2} for different
samples as well as for the disorder average.
With $2\,500$ MCS, the accuracy of the results for a given sample is not
perfect, of course, but the precision of the average over disorder is quite
good on the other hand. The disorder average procedure has been investigated
by computing the susceptibility $\chi_j$ for different samples, $1\leq j\leq
N_s$, where $N_s$ is the total number of samples (Fig.~\ref{Fig3}). We can see
that the dispersion of the values of $\chi$ is not very large because the
fluctuations in the average value disappear after a few hundreds
realisations.

      \begin{table}
       \caption{Evolution of the susceptibility with the number of Monte Carlo sweeps per spin
for different samples, $\chi_j$, and the average value (with $2\ 500$ samples) at $L=96$, $p=0.7$.
        \label{Tab2}}
\vspace*{0.1cm}
        \begin{tabular}{|l|l|l|l|l|l|l|} \hline
$\#\ {\rm MCS}\ $  & $\ \ \ \chi_1\ $  & $\ \ \ \chi_2\ $ & $\ \ \ \chi_3\ $ &
$\ \ \ \chi_4\ $ & $\ \ \ \chi_5\ $ & $\ \ \ \bar\chi\ $ \\ \hline
$\quad\ 100$ & $\ 1\,268\ $ & $\quad 720\ $ & $\ 1\,141\ $ & $\quad 939\ $ & $\quad 833\
$ & $\ 1\,058\ $     \\
$\quad\ 500$ & $\ 1\,272\ $ & $\ 1\,520\ $ & $\ 1\,223\ $ & $\ 1\,029\ $ & $\quad
953\ $ & $\ 1\,210\ $      \\
$\ \ 1\,000$ & $\ 1\,262\ $ & $\ 1\,544\ $ & $\ 1\,205\ $ & $\ 1\,068\ $ & $\quad
911\ $ & $\ 1\,219\ $     \\
$\ \ 1\,500$ & $\ 1\,282\ $ & $\ 1\,433\ $ & $\ 1\,277\ $ & $\ 1\,047\ $ & $\quad
915\ $ & $\ 1\,227\ $    \\
$\ \ 2\,000$ & $\ 1\,332\ $ & $\ 1\,441\ $ & $\ 1\,221\ $ & $\ 1\,073\ $ & $\quad
917\ $ & $\ 1\,235\ $     \\
$\ \ 2\,500$ & $\ 1\,358\ $ & $\ 1\,484\ $ & $\ 1\,234\ $ & $\ 1\,012\ $ & $\
1\,014\ $ & $\ 1\,234\ $      \\ 
\hline         
\end{tabular}        
\end{table}

\begin{figure}
\epsfysize=6.1cm
\begin{center}
\mbox{\epsfbox{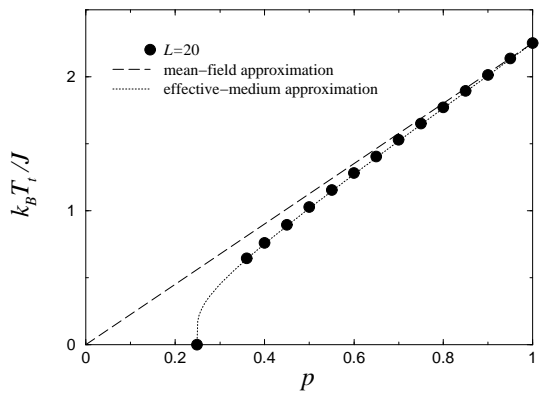}}
\end{center}\vskip 0cm
\caption{Phase diagram of the 3D bond-diluted Ising model compared with the 
mean-field and effective-medium approximations.}
\label{Fig4}
\end{figure}

The phase diagram obtained from the location of the maxima of the susceptibility for the largest
lattice size as a function of the concentration of magnetic bonds is shown in Fig.~\ref{Fig4}.
The simple mean-field transition temperature is drawn for comparison: it gives a linear
behaviour as a function of the bond concentration $p$ and we can check that this approximation
holds only in the low-dilution regime, $p>0.8$. On the other hand, the effective-medium
approximation~\cite{Turban1980} gives very good agreement with the simulated transition line.
Treating only a single bond in an effective medium leads to the following relation for the
critical coupling:
\begin{equation}
K_c (p)=\ln{{(1-p_c )\e^{K_c (1)}-(1-p)\over p-p_c}}.
\label{eq.3}
\end{equation}
This relation is exact in the vicinity of both the pure system and the percolation threshold.

\section{Competition between the fixed points}

The main problem encountered in previous studies of the disordered Ising model was the question
of measuring effective or asymptotic exponents. Although the change of universality class
should happen theoretically for an arbitrarily low disorder, it can be very difficult to measure
the new critical exponents because the asymptotic behaviour cannot always be reached
practically.
Another difficulty comes from the vicinity of the ratios $\gamma/\nu$ and $\beta/\nu$ in the
pure and disordered universality classes. Indeed, these values for the pure model
are \cite{Guida1998}:
$$\gamma/\nu=1.966(6),\ \beta/\nu=0.517(3),\ \nu=0.6304(13),$$
and for the disordered Ising model \cite{Ballesteros1998}:
$$\gamma/\nu=1.96(3),\ \beta/\nu=0.519(8),\ \nu=0.6837(53).$$

\begin{figure}
\epsfysize=6.cm
\begin{center}
\mbox{\epsfbox{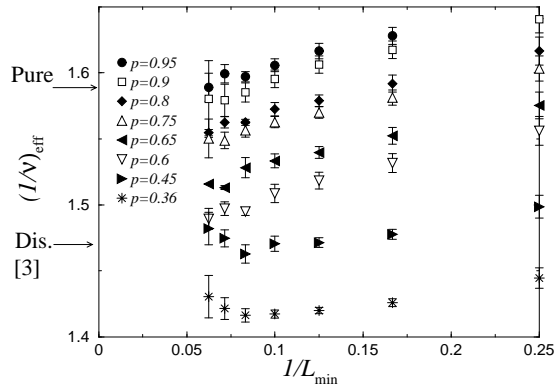}}
\end{center}\vskip 0cm
\caption{Effective exponents $(1/\nu)_{{\rm eff}}$ as a function of $1/L_{{\rm min}}$ for $p=0.95,\ 0.9,\ 0.8,
\ 0.75,\ 0.65,\ 0.6,\ 0.45$ and $0.36$. The error bars correspond to the standard deviations of
the power-law fits. The arrows indicate the values of $1/\nu$ for
the pure model \cite{Guida1998} and the site-diluted one \cite{Ballesteros1998}.}
\label{Fig5}
\end{figure}

Thus, from standard finite-size scaling techniques, the critical exponent $\nu$ only will allow
us to discriminate between the two fixed points. This exponent can be evaluated from the
finite-size scaling of the derivative of the magnetisation versus the temperature which is
expected to behave as ${\d\ln m\over \d K}\sim L^{1/\nu}$.
From this power-law behaviour, we have extracted the effective size-dependent exponent
$(1/\nu)_{{\rm eff}}$ which is plotted against
$1/L_{{\rm min}}$ for different bond concentrations $p$ in Fig.~\ref{Fig5} where $L_{{\rm min}}$ is the
smallest lattice size used in the fits.

We clearly see that in the regime of low dilution ($p$ close to $1$), the system is influenced
by the pure fixed point. On the other hand, when the bond concentration is small, the vicinity
of the percolation fixed point induces a decrease of $1/\nu$ below its expected disordered
value. Indeed, the percolation fixed point is characterized by
$1/\nu\sim 1.12$ \cite{Lorenz1998}.


\section{Conclusion}

We have presented, from a numerical study, the influence of bond dilution on the critical
properties of the 3D Ising model. The universality class of the disordered model is modified
by disorder but its precise characterization is difficult because of the competition between
the different fixed points which induce crossover effects, even for relatively
large lattice sizes. The next step is to accurately locate the best dilution
which minimizes these crossover effects.

We gratefully acknowledge financial support by the DAAD and EGIDE through the PROCOPE
exchange programme. C.C. thanks the EU network ``{\sl Discrete Random
Geometries: from solid state physics to quantum gravity}" for a postdoctoral
grant. This work was supported by the computer-time grants 2000007 of the
Centre de Ressources Informatiques de Haute-Normandie (CRIHAN) and hlz061 of
NIC, J\"ulich.


\begin{thebibliography}{00}



\bibitem{Harris74} A.B. Harris, J. Phys. C 7 (1974) 1671.

\bibitem{Pelissetto2000} A. Pelissetto and E. Vicari, Phys. Rev. B 62 
(2000) 6393.

\bibitem{Ballesteros1998} H.G. Ballesteros, L.A. Fern\'andez, V.
	Mart\'\i n-Mayor, A. Mu\~noz Sudupe, G. Parisi, and J.J. 
	Ruiz-Lorenzo, Phys. Rev. B 58 (1998) 2740.

\bibitem{Folk2001} R. Folk, Y. Holovatch, and T. Yavors'kii, 
e-print cond-mat/0106468.

\bibitem{Guida1998} R. Guida and J. Zinn-Justin, J. Phys. A 31 (1998) 8103.

\bibitem{Turban1980} L. Turban, Phys. Lett. 75A (1980) 307.

\bibitem{Lorenz1998} C.D. Lorenz and R.M. Ziff, Phys. Rev. E 57 (1998) 230.

\end{thebibliography}
\end{document}